\documentclass[english]{jetpl}

\twocolumn

\usepackage{amsmath,amssymb,color,pstricks,bm,alltt}

\usepackage{graphicx}

\begin{document}

\lat

\title{On Planckian limit for inelastic relaxation in metals}

\rtitle{On Planckian limit for inelastic relaxation in metals}

\sodtitle{On Planckian limit for inelastic relaxation i metals}

\author{M.\ V.\ Sadovskii\thanks{E-mail: sadovski@iep.uran.ru},
}

\rauthor{M.\ V.\ Sadovskii}

\sodauthor{M.\ V.\ Sadovskii}

\address{Institute for Electrophysics RAS, Ural Branch, Amundsen str. 106, 
Ekaterinburg 620016, Russia
}


\abstract{We consider the simplest model for $T$ -- linear growth of
resistivity in metals. It is shown that the so called ``Planckian'' limit for
the temperature dependent relaxation rate of electrons follows from a certain
procedure for representation of experimental data on resistivity and, in this
sense, is a kind of delusion.
}

\PACS{72.10.Di, 71.15.Cz, 72.15.Lh}

\maketitle

Linear with temperature growth of electrical resistivity in cuprates and some other
correlated systems in a wide region from very low to pretty high temperatures for 
many years remains one of the major puzzles of the physics of high -- temperature
superconductors. In recent years, a number of interesting papers appeared \cite{Bruin,Legros}, 
where by the analysis of experiments on rather wide range of compounds, it was shown 
that in the $T$ -- linear region of resistivity growth, the scattering rate of electrons
(inverse relaxation time) with rather high accuracy is described as
 $\Gamma=\frac{1}{\tau}=\alpha \frac{k_BT}{\hbar}$, where $\alpha\sim 1$ and is weakly
depending on the choice of the material. In particular, for systems being close to a
quantum critical point (on the phase diagram of cuprates and some other similar systems)
$\alpha$ belongs to the interval 0.7 -- 1.1. More so, the similar dependence is describes
rather well the data for a number of usual metals (like Cu, Au, Al, Ag, Pb, Nb etc.) 
in the region of $T$ -- linear growth of resistivity (which is usually realized at
temperatures $T>\Theta_D/5$, where $\Theta_D$ is Debye temperature). 
in this case $\alpha$ covers significantly wider interval from от 0.7 to 2.7 \cite{Bruin,Legros}. 
In connection with these (and some similar) results the notion of the universal (independent of
interaction strength) ``Planckian'' {\em upper} limit of scattering rate was introduced  $\frac{1}{\tau_P}
=\Gamma_P=\frac{k_BT}{\hbar}$ \cite{Zaanen}. To explain this temperature behavior of
resistivity for such different systems, also starting from very low temperatures, up to now
a number of relatively complicated theoretical models were proposed \cite{Khod,Khodel,Sachd,Vol}, 
including some rather exotic, based on the analogies taken from black hole physics, cosmology
and superstring theory (e.g. see Refs. \cite{Zan,Hart,Herz,Hartn}). In usual metals the
temperature dependence of resistivity (conductivity) is almost completely related to
inelastic scattering of electrons by phonons. In usual metals at high enough temperatures
$T>\Theta_D/5$ it dominates and leads to $T$ -- linear growth of resistivity:
\begin{equation}
\rho(T)-\rho_0=AT
\label{AT}
\end{equation}
where $\rho_0$ is the residual resistivity at $T=0$ due to the scattering by random impurities.

In terms of conductivity we may write the simple Drude expression:
\begin{equation}
\sigma(T)=\sigma_0+\sigma(T)
\label{Drude}
\end{equation}
where $\sigma_0$ is the residual conductivity at $T=0$ and
\begin{eqnarray}
\sigma(T)=\frac{ne^2}{m}\tau(T)=\frac{ne^2}{m}\Gamma^{-1}(T)
\label{sigma}
\end{eqnarray}
Here and below $m$ is understood to be the {\em band} effective mass,
while $\Gamma(T)=\frac{1}{\tau(T)}$ is the temperature dependent relaxation 
(scattering) rate due to inelastic scattering of electrons by phonons, which
grows linearly with temperature for $T>0.2\Theta_D$. Correspondingly we get
the resistivity as:
\begin{equation}
\rho(T)-\rho_0=\frac{m}{ne^2}\Gamma(T)
\label{sigmainv}
\end{equation}

The concept of ``Planckian'' relaxation rate can be introduced via elementary
estimates \cite{Hart}. At $T>0$ the processes of inelastic scattering appear
due to different interactions (electron -- phonon, spin fluctuations,
quantum fluctuation of arbitrary origin). In particular, these processes are
responsible for thermodynamic equilibrium of electronic subsystem leading to
Fermi distribution. Conductivity of a metal (degenerate case) is determined by
electrons distributed in a layer of the width $\sim k_BT$ around the Fermi
level (chemical potential).

Let us perform an elementary estimates using the energy -- time uncertainty
relation:
\begin{equation}
\Delta E\tau > \hbar
\label{Heis}
\end{equation}
where $\tau$ is the lifetime of a quantum state and $\Delta E$ is it energy
uncertainty. Naturally, in our case $\tau=\tau(T)$, while $\Delta E\sim k_BT$, 
which immediately leads to an estimate
\begin{equation}
\Gamma(T)=\frac{1}{\tau(T)}<\alpha\frac{k_BT}{\hbar}\equiv\alpha\Gamma_P=
\frac{\alpha}{\tau_P}
\label{Planck}
\end{equation}
where $\alpha\sim 1$. We conclude that according to such elementary estimate
the ``Planckian'' relaxation rate determines precisely the {\em upper} limit
for resistivity due to inelastic scatterings:
\begin{equation}
\rho(T)-\rho_0=\frac{m}{ne^2}\Gamma(T)<\frac{m}{ne^2}\alpha\frac{k_BT}{\hbar}\equiv \alpha\rho_P(T)
\label{R_Planck}
\end{equation}
However, it is obvious that this estimate is of rather speculative nature for
the system of many interacting particles.

Consider the following Hamiltonian for interaction of metallic electrons with
arbitrary quantum Bose -- type fluctuations\footnote{In the following we assume
$\hbar=k_B$=1}:
\begin{equation}
H_{int}=\frac{1}{N}\sum_{\bf pq}g_{\bf q}a^+_{\bf p+q}a_{\bf p}\rho_{\bf q}
\label{Hami}
\end{equation}
Here we use the standard notations for creation and annihilation operators of
electrons, $\rho_{\bf q}$ is the quantum fluctuation operator of ``any kind'' 
(e.g. ion density in a lattice), $g_{\bf q}$ is the appropriate coupling
constant (matrix element of interaction potential) \cite{Pin,PinNoz}. 
Let us introduce the appropriate (Matsubara) Green's function as:
\begin{equation}
F({\bf q},\tau)=-<T_{\tau}\rho_{\bf q}(\tau)\rho^+_{\bf q}(0)>
\label{GrFluc}
\end{equation}
For this function we can write down the standard (Bose) spectral
representation \cite{AGD}:
\begin{equation}
F({\bf q},i\omega_m)=\int_{-\infty}^{\infty}d\omega\frac{A({\bf q},\omega)}
{i\omega_m-\omega}
\label{spectrrep}
\end{equation}
where $\omega_m=2\pi mT$ and spectral density is defined as:
\begin{equation}
A({\bf q},\omega)=Z^{-1}\sum_{mn}e^{-\frac{E_n}{T}}|(\rho_{\bf q})_{nm}|^2
\left(1-e^{-\frac{\omega_{mn}}{T}}\right)\delta(\omega-\omega_{mn})
\label{specdens}
\end{equation}
where $\omega_{mn}=E_m-E_n$, $(\rho_{\bf q})_{nm}=<n|\rho_{\bf q}|m>=(\rho^+_{\bf q})_{mn}$.

Dynamic structure factors of fluctuations is defined as \cite{Pin,PinNoz}:
\begin{equation}
S({\bf q},\omega)=Z^{-1}\sum_{mn}e^{-\frac{E_n}{T}}|(\rho_{\bf q})_{nm}|^2
\delta(\omega-\omega_{mn})
\label{dinstr}
\end{equation}
Comparing (\ref{specdens}) and (\ref{dinstr}) we obtain:
\begin{equation}
A({\bf q},\omega)=S({\bf q},\omega)\left[1-e^{-\frac{\omega}{T}}\right]
\label{AS}
\end{equation}

\begin{figure}
\includegraphics[clip=true,width=0.5\textwidth]{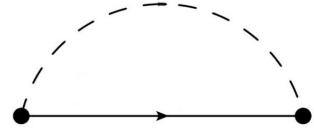}
\caption{Diagram of the second order for electron self -- energy.
Dashed line -- Green's function of quantum fluctuations $F$,
continuous line -- electron Green's function $G$.}
\label{SE}
\end{figure}

Electronic Greens's function in Matsubara representation is written as:
\begin{equation}
G(\varepsilon_n,{\bf p})=\frac{1}{i\varepsilon_n-\xi_{\bf p}-\Sigma(\varepsilon_n,{\bf p})}
\label{GrF}
\end{equation}
where $\varepsilon_n=(2n+1)\pi T$, $\xi_{\bf p}$ is electronic spectrum counted
from the chemical potential. Assuming the validity of Migdal's theorem \cite{Diagram}
we can take the electron self -- energy in the simplest approximation, shown
in Fig. \ref{SE}:
\begin{eqnarray}
\Sigma(\varepsilon_n,{\bf p})=
\frac{T}{N}\sum_{\bf q}g^2_{\bf q}\sum_{m}F({\bf q},i\omega_m)
G(\varepsilon_n+\omega_m,{\bf p+q})\nonumber\\
=\frac{T}{N}\sum_{\bf q}g^2_{\bf q}\sum_{m}\int_{-\infty}^{\infty}d\omega\frac{S({\bf q},\omega)}
{i\omega_m-\omega}
\left(1-e^{-\frac{\omega}{T}}\right)\nonumber\\
\frac{1}{i\varepsilon_n+i\omega_m-\xi_{\bf p+q}}\nonumber\\
\label{SEFG}
\end{eqnarray}
Consider now the case when the average frequency of fluctuations
$\langle\Omega\rangle$ is significantly lower than the temperature $T$. 
Then in Eq. (\ref{SEFG}) we can limit ourselves only by term with
$m=0$ and thus to the picture of {\em quasielastic} scattering by fluctuations: 
\begin{eqnarray}
\Sigma(\varepsilon_n,{\bf p})=\frac{1}{N}\sum_{\bf q}g^2_{\bf q}\int_{-\infty}^{\infty}d\omega
\frac{S({\bf q},\omega)}
{i\varepsilon_n-\xi_{\bf p+q}}=\nonumber\\
=\sum_{\bf q}g^2_{\bf q}S({\bf q})\frac{1}
{i\varepsilon_n-\xi_{\bf p+q}}
\label{Sen}
\end{eqnarray}
where we have introduced the structure factor of fluctuations as \cite{PinNoz}:
\begin{equation}
S({\bf q})=\frac{1}{N}\int_{-\infty}^{\infty}d\omega S({\bf q},\omega)
\label{StrFac}
\end{equation}
In fact this is a direct analog the well-known Ziman -- Edwards approximation
in the theory of liquid metals. The case of $S({\bf q})=const$ corresponds to
chaotic distribution of static scattering centers \cite{Diagram}.

Fluctuation operator $\rho_{\bf q}$ for the case of some collective mode can be
expressed via (Boson) annihilation and creation operators for corresponding
quanta (e.g. phonons) \cite{PinNoz}:
\begin{equation}
\rho_{\bf q}=\frac{1}{\sqrt 2}\left(b^+_{\bf q}+b_{-\bf q}\right)
\label{phons}
\end{equation}
Then we have:
\begin{eqnarray}
S({\bf q},\omega)=Z^{-1}\sum_m e^{-\beta E_m}\left[<m|b_{\bf q}b^+_{\bf q}|m>
\delta(\omega-\omega_{\bf q})+\right.\nonumber\\
\left.+<m|b^+_{-\bf q}b_{-\bf q}|m>\delta(\omega+\omega_{\bf q})\right]\nonumber\\
\label{Sqw}
\end{eqnarray}
where $\omega_{\bf q}$ is the spectrum of corresponding excitations.
Introducing the usual Bose distribution:
\begin{equation}
n_{\bf q}=Z^{-1}\sum_m e^{-\beta E_m}<m|b^+_{\bf q}b_{\bf q}|m>=\frac{1}{e^{\beta\omega_{\bf q}}-1}
\label{bose}
\end{equation}
we get \cite{PinNoz}
\begin{eqnarray}
S({\bf q},\omega)=\left[(n_{\bf q}+1)\delta(\omega-\omega_{\bf q})
+n_{\bf q}\delta(\omega+\omega_{\bf q})\right]=\nonumber\\
=\delta(\omega-\omega_{\bf q})+n_{\bf q}\left[\delta(\omega-\omega_{\bf q})+\delta(\omega+\omega_{\bf q})\right]
\label{Swqq}
\end{eqnarray}
In case of $T\gg\omega_{\bf q}$, we have:
\begin{equation}
n_{\bf q}=\frac{T}{\omega_{\bf q}}
\label{qclass}
\end{equation}
and correspondingly:
\begin{equation}
S({\bf q})=\frac{2T}{\omega_{\bf q}}
\end{equation}
i.e. we obtain the structure factor linear in $T$ and its momentum dependence
is determined simply by excitation spectrum of the appropriate collective mode
(fluctuation). Then:
\begin{equation}
\Sigma(\varepsilon_n,{\bf p})=T\sum_{\bf q} \frac{2 g^2}{\omega_{\bf q}}\frac{1}{i\varepsilon_n-\xi_{\bf p+q}}
\label{SEel}
\end{equation}
To simplify the model further let us assume the spectrum of fluctuations to be
dispersionless (like Einstein phonon or optical phonon with weak dispersion)
$\omega_{\bf q}=\Omega_0$. Then performing all calculations similarly to the
problem of an electron in the field of random impurities \cite{Diagram}, we get:
\begin{equation}
\Sigma(\varepsilon_n,{\bf p})=-i\pi sign \varepsilon_n\frac{2g_0^2}{\Omega_0}N(0)T
\label{SiGG}
\end{equation}
where $N(0)$ is the density of states at the Fermi level.
Correspondingly, the scattering rate (damping) is written as:
\begin{equation}
\frac{\Gamma(T)}{2}=\pi\frac{2g_0^2}{\Omega_0}N(0)T=\pi\lambda_0 T
\label{GamT}
\end{equation}
where we introduced in a standard way the dimensionless coupling constant:
\begin{equation}
\lambda_0=\frac{2g^2_0N(0)}{\Omega_0}
\label{Lamb_st}
\end{equation}
Now the electronic Green's function takes the usual form \cite{Diagram}:
\begin{equation}
G(i\varepsilon_n,{\bf p})=\frac{1}{i\varepsilon_n-\varepsilon_{\bf p}+\frac{i}{2}
\Gamma(T) sign\varepsilon_n}
\label{MatGrP}
\end{equation}
where are no renormalization factors of any kind (the residue at the pole of Green's
function $Z=1$), which is natural for temperatures much exceeding the frequencies of
quantum fluctuations.

After the standard calculations \cite{Diagram} we obtain the resistivity as:
\begin{equation}
\rho(T)=\frac{m}{ne^2}\Gamma(T)=2\pi\lambda_0\rho_{P}(T)
\label{ResistT}
\end{equation}
which in essence just coincide with high -- temperature limit of Eliashberg -- McMillan
theory \cite{Savr} . The constant $\alpha$ used in writing down the ``Planckian'' 
relaxation as (\ref{Planck}) is expressed via the the parameters of the theory as;
\begin{equation}
\alpha=2\pi\lambda_0
\label{alP}
\end{equation}
Naturally, it is not universal and is just proportional to the coupling constant.

All this is known actually for a long time and trivially explains the $T$ -- linear
growth of resistivity in accordance with many experiments. To make such resistivity
growth starting from low temperatures it is sufficient to demand that $\Omega_0\ll T$. 
In the vicinity of the quantum critical point (of any nature) we can expect the
typical ``softening'' of fluctuation modes like \cite{Sachdev}:
\begin{equation}
\Omega_0\sim |x-x_c|^{z\nu}
\label{softm}
\end{equation}
where $x$, for example, may denote the carrier concentration close to the critical $x_c$,
while $\nu$ and $z$ are the standard critical exponents of the theory of quantum phase
transitions, determining the critical behavior of characteristic lengths:
\begin{equation}
\xi\sim|x-x_c|^{-\nu},\ \xi_{\tau}\sim|x-x_c|^{-z\nu}
\label{lengths}
\end{equation}
where $\tau$ is the imaginary (Matsubara) time, so that above we may just assume
$\Omega_0\sim\xi_{\tau}^{-1}$. This may be responsible for $T$ -- linear behavior
in systems like cuprates.

We can avoid the explicit introduction of phonons (or any other quasiparticles related to
fluctuations). From Eq. (\ref{AS}) for $\omega\ll T$ we have:
\begin{equation}
A({\bf q},\omega)\approx \frac{\omega}{T}S({\bf q},\omega)
\label{Aq}
\end{equation}
or
\begin{equation}
S({\bf q},\omega)\approx \frac{T}{\omega}A({\bf q},\omega)
\label{Sq}
\end{equation}
Substituting this expression into Eq. (\ref{Sen}) we get the following expression for
self -- energy:
\begin{equation}
\Sigma(\varepsilon_n,{\bf p})=
\frac{T}{N}\sum_{\bf p'}g^2_{\bf pp'}\int_{-\infty}^{\infty}\frac{d\omega}{\omega}
\frac{A({\bf p-p'},\omega)}
{i\varepsilon_n-\xi_{\bf p'}}
\label{Senr}
\end{equation}
where everything is determined by the spectral density of fluctuations $A({\bf q},\omega)$, 
which is not necessarily of quasiparticle form. Obviously, for the simplest model
with $A({\bf q},\omega)=\delta(\omega-\Omega_0)$ (Einstein model of fluctuations)
from Eq. (\ref{Senr}) we immediately obtain the previous results of Eqs.
(\ref{SiGG}) -- (\ref{Lamb_st}).

Further, let us assume that fluctuations scatter electrons in some pretty narrow
layer around the Fermi surface with width determined by their characteristic
frequencies ($\langle\Omega\rangle\ll T$). Then, in the spirit of Eliashberg --
McMillan theory we can introduce the self -- energy averaged over momenta 
at the Fermi surface:
\begin{equation}
\Sigma(\varepsilon_n)=\frac{1}{N(0)}\sum_{\bf p}\delta{(\xi_{\bf p})}
\Sigma(\varepsilon_n,{\bf p}),
\label{SigMac}
\end{equation}
and also an effective interaction averaged over the initial and final momenta at 
the Fermi surface:
\begin{eqnarray}
g_{\bf pp'}^2A({\bf p-p',\omega})\Longrightarrow\nonumber\\
\frac{1}{N(0)}\sum_{\bf p}\frac{1}{N(0)}\sum_{\bf p'}g_{\bf pp'}^2A({\bf p-p',\omega})
\delta(\xi_{\bf p})\delta({\xi_{\bf p'}})\nonumber\\
\equiv\frac{1}{N(0)}\alpha^2(\omega)F(\omega)
\label{McMEavr}
\end{eqnarray}
where
\begin{equation}
F(\omega)=\sum_{\bf q}A({\bf q,\omega})
\label{flDOS}
\end{equation}
is the density of states of fluctuations. 
Then for (\ref{SigMac}) from (\ref{Senr}) we get:
\begin{eqnarray}
\Sigma(\varepsilon_n)=\frac{T}{N(0)}\int_{-\infty}^{\infty}
\frac{d\omega}{\omega}\alpha^2(\omega)F(\omega)N(0)\int_{-\infty}^{\infty} d\xi\frac{1}
{i\varepsilon_n-\xi}\nonumber\\
=-i\pi sign\varepsilon_n T\int_{-\infty}^{\infty}\frac{d\omega}{\omega}
\alpha^2(\omega)F(\omega)\nonumber\\
=-i\pi sign\varepsilon_n\lambda T 
\equiv -i\frac{\Gamma(T)}{2}sign\varepsilon_n\nonumber\\
\label{SE-MCM}
\end{eqnarray}
where we have introduced the dimensionless coupling constant similar to that
in Eliashberg -- McMillan theory:
\begin{equation}
\lambda=2\int_{0}^{\infty}\frac{d\omega}{\omega}\alpha^2(\omega)F(\omega)
\label{lambdaMcM}
\end{equation}
which is in fact determined by (averaged according to (\ref{McMEavr}))
the spectral density of fluctuations $A({\bf q},\omega)$.

Then we obtain:
\begin{equation}
\Gamma(T)=2\pi\lambda T
\label{Plkrel}
\end{equation}
which is formally of the same form as (\ref{GamT}) and immediately leads to
(\ref{ResistT}). 

In Refs. \cite{Bruin,Legros} experimental data on resistivity were fitted to
standard Drude expression (\ref{sigmainv}), where the effective mass was
determined from low temperature measurements (electronic specific heat and
oscillation effects in high magnetic fields) which is actually related to
band structure effective mass by a simple replacement
$m\to\tilde m=m(1+\lambda)$, which explicitly takes into account renormalization due to
phonons. The deficiency of such approach was already noted in Ref. \cite{Varma}. 
Let us show that precisely this kind of representation of data creates a
{\em delusion} of universal ``Planckian'' relaxation. In fact, the expression
(\ref{ResistT}) for {\em high -- temperature} limit of resistivity can be rewritten as:
\begin{equation}
\rho(T)=\frac{m(1+\lambda)}{ne^2}\frac{\Gamma(T)}{1+\lambda}=\frac{\tilde m}{ne^2}\tilde\Gamma(T)
\label{ResisT}
\end{equation}
where
\begin{equation}
\tilde\Gamma(T)=2\pi\frac{\lambda}{1+\lambda}T
\label{tildaGamma}
\end{equation}
which for $\lambda\gg 1$ reduces to:
\begin{equation}
\tilde\Gamma(T)=2\pi T
\label{PlanckGamma}
\end{equation}
and simply imitates the universal ``Planckian'' behavior of relaxation rate 
(\ref{Planck}) with $\alpha=2\pi$, which is independent of coupling constant of electrons with fluctuations
(phonons). The replacement $m\to\tilde m=m(1+\lambda)$ in Eq. (\ref{ResisT}) is correct despite the fact, that
here we are dealing with high -- temperature limit as fitting the experimental data in Refs. \cite{Bruin,Legros} 
was performed using the effective mass $\tilde m$, obtained from {\em low temperature} measurements, which contains
renormalization effects. For quantitative estimates it is also quite important to take into account mass
renormalization due to interelectron interactions. Correspondingly, Eq. (\ref{tildaGamma}) should be rewritten as:
\begin{equation}
\tilde\Gamma(T)=2\pi\frac{\lambda}{1+\lambda+\lambda_{ee}}T
\label{tildaGmm}
\end{equation}
where $\lambda_{ee}$ is dimensionless parameter, determining mass renormalization due to
electron -- electron interactions. In Landau -- Silin theory of Fermi liquids 
$\lambda_{ee}=\frac{F^s_1}{3}$, where $F^s_1$ is the appropriate coefficient in Landau
function expansion \cite{PinNoz}.  Direct DMFT calculations for the Hubbard model produce
the values of renormalization factor $Z=(1+\lambda_{ee})^{-1}$ in metallic phase monotonously changing with
Hubbard interaction $U$ in the interval between 1 and 0 \cite{Kotl}. Thus, for rough estimates
for typical metal we can safely take $\lambda_{ee}\sim 1$. 
Then:
\begin{equation}
\alpha=\frac{2\pi\lambda}{1+\lambda+\lambda_{ee}}
\label{alpgener}
\end{equation}
Then the interval of $\alpha=0.7-2.7$ \cite{Bruin,Legros} for $\lambda_{ee}=1$
corresponds to $\lambda=0.25-1.5$, which seems quite reasonable.
For example for Al we have the calculated value $\lambda=0.44$ \cite{Savr}, which 
immediately gives $\alpha=1.03$  in nice correspondence with ``experimental''
value $\alpha=1$ from Ref. \cite{Bruin}. For Pb, taking $\lambda=1.68$ \cite{Savr} we
get $\alpha\sim 2.86$ in reasonable agreement with $\alpha=2.7$ \cite{Bruin}. 
Similarly, for Nb we have $\lambda=1.26$ \cite{Savr} and $\alpha\sim 2.42$, also in
good agreement with $\alpha=2.3$ of Ref.  \cite{Bruin}. In general, taking into account
the roughness of our estimate of $\lambda_{ee}\sim 1$ this agreement seems rather 
convincing\footnote{We neglect rather insignificant \cite{Savr} difference between
$\lambda$ and $\lambda_{tr}$.}.

Thus, the  ``experimentally'' observed universal ``Planckian'' relaxation rate in metals, 
independent of interaction strength, is nothing more than {\em delusion}, connected with 
the procedure used in Refs. \cite{Bruin,Legros} to represent the experimental data.
At low temperatures  ($T<\langle\Omega\rangle$) Green's function takes the form:
\begin{equation}
G(\varepsilon_n,{\bf p})=\frac{Z}{i\varepsilon_n-Z\xi_{\bf p}+\frac{i}{2}Z\Gamma(T)sign \varepsilon_n}
\label{G_dressed}
\end{equation}
where the renormalization factor $Z<1$ is assumed for simplicity a constant. 
The term $Z\Gamma(T)=\tilde\Gamma(T)$ in the denominator describes quasiparticle damping for
which it may seem we have the ``universal'' high -- temperature limit of Eq. (\ref{PlanckGamma}). 
However, it is wrong -- at high temperatures ($T>\langle\Omega\rangle$) the renormalization factor
$Z\to 1$, as can be seen from our results above. Also for the low temperatures, when 
$Z<1$, the term $Z\xi_{\bf p}$ in the denominator of (\ref{G_dressed}) describes the renormalized
spectrum of electrons with mass $\tilde m=m(1+\lambda)$, so that electron velocity at the Fermi
surface $v_F=p_F/m\to \tilde v_F=p_F/\tilde m=v_F/(1+\lambda)$. Renormalization of damping
corresponds to renormalization of mean free time $\tilde\Gamma^{-1}=\Gamma^{-1}(1+\lambda)$.
Now we see, that the mean free path is not renormalized: $l=\tilde v_F\tilde\Gamma^{-1}
=v_F\Gamma^{-1}$ and renormalization due to many particle effects, important at low temperatures,
actually do not affect conductivity (resistivity) at all \cite{Grim}.
In fact, this follows from the general Ward identity \cite{Hein}.

The author is grateful to E.Z. Kuchinskii for numerous useful discussions.
This work was partly supported by RFBR grant No. 20-02-00011.


\end{document}